\documentstyle[prl,aps,epsf,preprint]{revtex}
\begin{document}
\draft

\title{Gas molecule adsorption in carbon nanotubes and nanotube bundles}

\author{Jijun Zhao $^a$$^*$, Alper Buldum $^a$, Jie Han $^b$, Jian Ping Lu $^a$ $^{\dagger}$}

\address{$^a$: Department of Physics and Astronomy, University of North Carolina at Chapel Hill, Chapel Hill, NC 27599 \\
$^b$: NASA Ames Research Center, Mail Stop T27A-1, Moffett Field, CA 94035}

\date{\today}
\maketitle

\begin{abstract}

We studied various gas molecules (NO$_2$, O$_2$, NH$_3$, N$_2$, CO$_2$, CH$_4$, H$_2$O, H$_2$, Ar) on single-walled carbon nanotubes (SWNTs) and bundles using first principles methods. The equilibrium position, adsorption energy, charge transfer, and electronic band structures are obtained for different kinds of SWNTs. Most molecules adsorb weakly on SWNTs and can be either charge donor or acceptor to the nanotubes. We find the gas adsorption on the bundle interstitial and groove sties is stronger than that on an individual tube. The electronic properties of SWNTs are sensitive to the adsorption of certain gases such as NO$_2$ and O$_2$. Charge transfer and gas-induced charge fluctuation might significantly affect the transport properties of SWNTs. Our theoretical results are consistent with recent experiments. 

\end{abstract}


\section{Introduction}

In recent years, carbon nanotubes have been intensively studied due to their importance as building block in nanotechnology. The special geometry and unique properties of carbon nanotube offer great potential applications, including nanoelectronic devices, energy storage, chemical probes and biosensors, field emission display, etc \cite{1,2,3,4}. Gas adsorption in carbon nanotubes and nanotube bundles is an important issue for both fundamental research and technique application of nanotubes. Considerable experimental and theoretical efforts have been devoted to the hydrogen storage in nanotube-based materials \cite{5,6,7,8,9}. The effect of gas environment on the electronic properties of carbon nanotube have recently attracted certain attention \cite{14,15,16,17,18,19,20}. Upon exposure to O$_2$, NO$_2$, or NH$_3$ gas, the electrical conductance of the semiconducting tubes are dramatically changed \cite{14,15}. NMR measurements reveal the effect of oxygen on density of state at Fermi level \cite{16}. The resistance of metallic nanotube bundle decreases as gas molecules are removed from the sample \cite{17}. In the field emission application of carbon nanotubes, the influence of various residual gases in vacuum chamber is a critical factor for the long term stability \cite{18,19}. Theoretically, it has been shown that the O$_2$ adsorption has significant effect on electronic properties of small semiconducting nanotubes \cite{20}. In this paper, we report first principles calculations on individual SWNTs and tube bundles with adsorption of variety of gas molecules including NO$_2$, O$_2$, NH$_3$, N$_2$, H$_2$O, Ar, etc. 

\section{Computational methods}

The self-consistent field (SCF) electronic structure calculations are performed based on density functional theory (DFT) with either localized basis (DMol) or plane-wave basis (CASTEP). The equilibrium geometry, adsorption energy and charge transfer are calculated by using DMol program \cite{21}. During the DMol all electron SCF calculations, a double numerical basis including $d$-polarization function (DND) are adopted. The density functional is treated by the local density approximation (LDA) with the exchange-correlation potential parameterized by Perdew and Wang \cite{22}. Geometry optimizations are performed with the Broyden-Fletcher-Goldfarb-Shanno (BFGS) algorithm with convergence criterion of 10$^{-3}$ a.u on the gradient and displacement, and 10$^{-5}$ a.u. on the total energy and electron density. More accurate electronic band structure and electron density are calculated by a SCF plane-wave pseudopotential technique (CASTEP) \cite{23}. The ion-electron interaction is modeled by Troullier-Martin norm-conserving nonlocal pseudopotential \cite{24}. The energy cutoff of plane wave basis is chosen up to 760 eV. 

To check the validity of present theoretical scheme, we have studied several well-known cases of gas molecule adsorption on the graphene sheet. In Table I, we compare our LDA results with those ``best estimated'' experimental data given in Ref.[25]. We find that both the equilibrium distance and adsorption energy are well reproduced by our present DFT scheme. 

\begin{table}
Table I. Equilibrium height ($h$, defined by the distance between molecule mass center and the center of six-member carbon ring on graphite), adsorption energy ($E_a$) of various molecules on the graphene sheet. LDA denotes current LDA calculations, exp. is the ``best estimated'' experimental data \cite{25}.
\begin{center}
\begin{tabular}{cccccc} 
                      &  CO$_2$ & CH$_4$  & N$_2$  &   H$_2$  &  Ar    \\ \hline
$h$$^{LDA}$ (\AA)     &   3.11  & 3.34    & 3.31   &   2.92   &  3.31  \\
$E_a$$^{LDA}$ (meV)   &   151   &  154    & 110    &   92     &  97    \\ \hline
$h$$^{exp}$ (\AA)     &   3.2   &  3.45   & 3.34   &   2.87   &  3.1   \\
$E_a$$^{exp}$ (meV)   &   178   &  126    & 104    &   42     &  99    \\
\end{tabular}
\end{center}
\end{table}

In this work, we studied both zigzag (10,0), (17,0) and armchair (5,5), (10,10) tubes. One-dimensional periodic boundary condition is applied along tube axis. For (10,0) and (17,0) zigzag tubes, we use one molecule per unit cell in tube axis direction. Using one molecule per two unit cells shows no significant difference. For (5,5) or (10,10) tube, due to the short unit cell length, reliable results is obtained by using one molecule per two unit cells. For individual SWNTs, we have carried out static calculations to obtain the binding curve (see Fig.2) and find the equilibrium tube-molecules distance for each system. Full geometrical minimizations are then performed to determine the optimal molecular orientations and distance. Different possible adsorption sites, T (top of an carbon atom), B (top of the center of the C-C bond), C (top of the center of carbon hexagon) have been considered in our calculation. Upon relaxation, we find the deformation of nanotube structure is relatively small and does not significantly modify the electronic properties. In addition to individual SWNTs, we have also investigate the gas adsorption in the SWNTs bundle. A lattice constant 16.8 \AA$~$ for the (10,10) tube bundle is chosen \cite{26}. 

\section{Results and discussions}

\subsection{Interaction between gas molecule and an individual SWNT}

Table II summarizes our results on the equilibrium tube-molecule distance, adsorption energy, and charge transfer for various molecules on (10,0), (17,0) and (5,5) SWNTs \cite{18}. In general, all these gas molecules are weakly binded to nanotube and the tube-molecule interaction can be identified as physisorption. Most molecules studied (with exception of NO$_2$ and O$_2$) are charge donors with small charge transfer (0.01 $\sim$ 0.035 electron per molecule) and weak binding ($\leq$ 0.2 eV). For O$_2$ and NO$_2$, both of which are charge acceptor, the charge transfer is not negligible. This is also reflected in their larger adsorption energies.

\newpage
\begin{table}
Table II. Equilibrium tube-molecule distance ($d$), adsorption energy ($E_a$) and charge transfer ($Q$) of various molecules on (10,0), (17,0) and (5,5) individual SWNTs \cite{27}. The optimal adsorption sites are given  in the table:: T (top of an carbon atom), B (top of the center of the C-C bond), C (top of the center of carbon hexagon). 
\begin{center}
\begin{tabular}{cccccccccc} 
          & \multicolumn{9}{c}{ (10,0) SWNT} \\ 
          & NO$_2$&  O$_2$ & H$_2$O & NH$_3$ & CH$_4$ & CO$_2$ & H$_2$ &  N$_2$ &  Ar   \\ \hline
$d$(\AA)  & 1.93  &  2.32  &  2.69  &  2.99  & 3.17   & 3.20   & 2.81  & 3.23   & 3.32  \\
$E_a$(meV)& 797   &  509   &  143   &  149   & 190    &  97    & 113   &  164   &  57   \\
$Q$(e)    & -0.061& -0.128 & 0.035  & 0.031  & 0.027  & 0.016  & 0.014 &  0.008 & 0.01   \\
Site      &  T    &    B   &   T    &   T    &    C   &   C    &   C   &   C    &   C   \\ \hline \ \\

          & \multicolumn{8}{c}{ (5,5) SWNT} \\
          & NO$_2$&  O$_2$ & H$_2$O & NH$_3$ & CH$_4$ & CO$_2$ & H$_2$ &  N$_2$ &  Ar   \\ \hline
$d$(\AA)  &  2.16 &  2.46  &  2.68  &  2.99  & 3.33   &  3.54  & 3.19  & 3.23   & 3.58  \\
$E_a$(meV)&  427  &  306   &  128   &  162   & 122    & 109    & 84    & 123    &  82   \\
$Q$(e)    & -0.071& -0.142 & 0.033  & 0.033  & 0.022  & 0.014  & 0.016 & 0.011  & 0.011 \\ 
Site      &  T    &    B   &   T    &   T    &    C   &   C    &   C   &   C    &   C   \\ \hline \ \\

          & \multicolumn{8}{c}{ (17,0) SWNT} \\
          & NO$_2$&  O$_2$ & H$_2$O & NH$_3$ & CH$_4$ & CO$_2$ & H$_2$ &  N$_2$ &  Ar   \\ \hline
$d$(\AA)  & 2.07  &  2.50  &  2.69  & 3.00   & 3.19   &  3.23  & 2.55  & 3.13   &  3.34 \\
$E_a$(meV)& 687   &  487   &  127   & 133    & 72     &  89    &  49   & 157    &  82   \\
$Q$(e)    & -0.089& -0.096 & 0.033  & 0.027  & 0.025  & 0.015  & 0.012 &  0.006 &  0.01   \\ 
Site      &  T    &    B   &   T    &   T    &    C   &   C    &   C   &   C    &   C   \\ 
\end{tabular}
\end{center}
\end{table}

\begin{figure}
\centerline{
\epsfxsize=4.5in \epsfbox{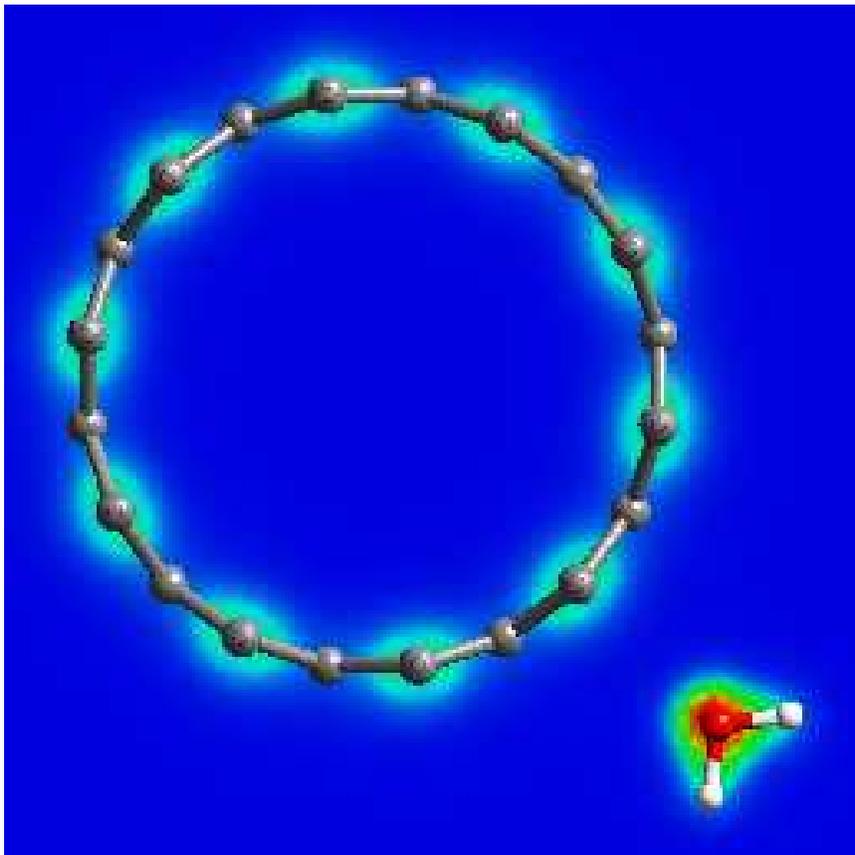}
}
\caption{Geometric structure and total electron density (slice at (100) direction) of H$_2$O attach to an individual (10,0) tube. No significant overlap of the electron density is found between molecule and tube.}
\end{figure}

To illustrate the tube-molecule interaction, Fig.1 shows the atomic structure and total charge density of valence electrons for a H$_2$O molecule adsorbed on (10,0) SWNT. No substantial electron density overlap is found in the region between gas molecule and nanotube, indicating that there is no chemical bond formed. This result agree well with a recent DMol calculation \cite{13}, which show that the water-tube interaction is weak without electric field and can be significantly enhanced by applied field. The binding curve for H$_2$O and NO$_2$ molecules adsorbed on (10,0), (17,0) and (5,5) SWNTs are shown in Fig.2. The tube-molecule interactions are comparable to the van der Waals-like interactions between these molecules and graphite surface \cite{25}. our results show that there is no clear dependence of adsorption on the tube size and chirality (Table II and Fig.2).

\begin{figure}
\centerline{
\epsfxsize=4.5in \epsfbox{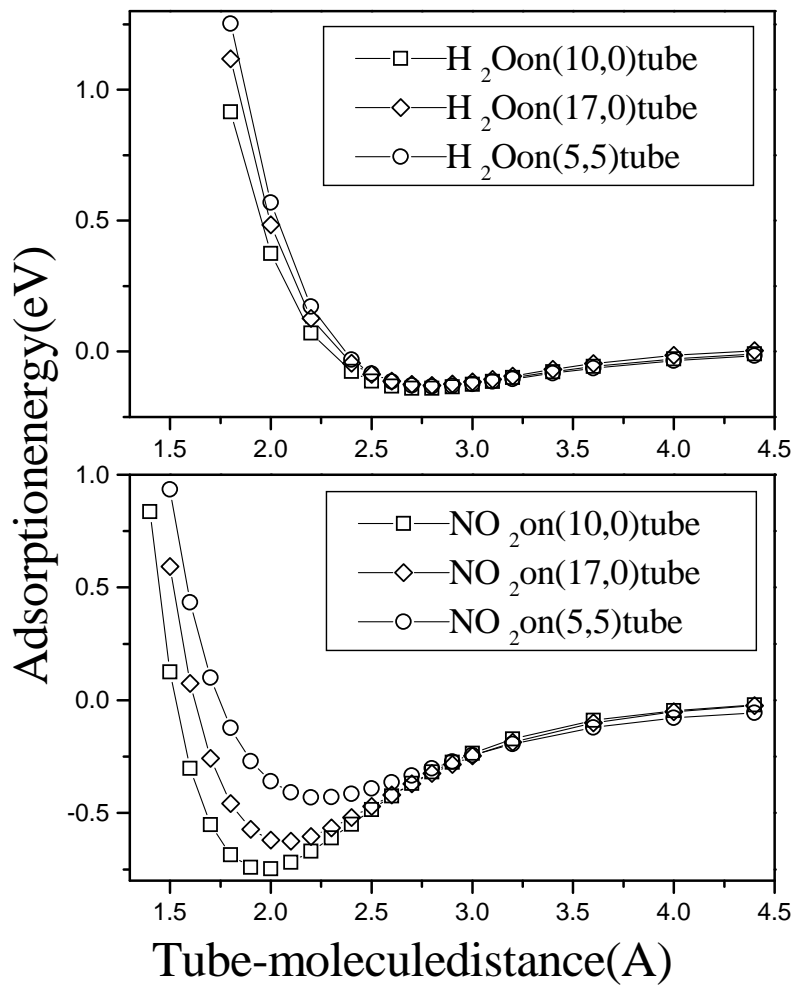}
}
\caption{Adsorption energy (eV) as functions of tube-molecule distance ($\AA$) (defined as the nearest distance between the molecule and the nanotube) for H$_2$O (above) and NO$_2$ (below) on (10,0), (17,0) and (5,5) tubes. }
\end{figure}

\subsection{Gas molecule in tube bundle and on bundle surface}

\begin{figure}
\centerline{
\epsfxsize=4.5in \epsfbox{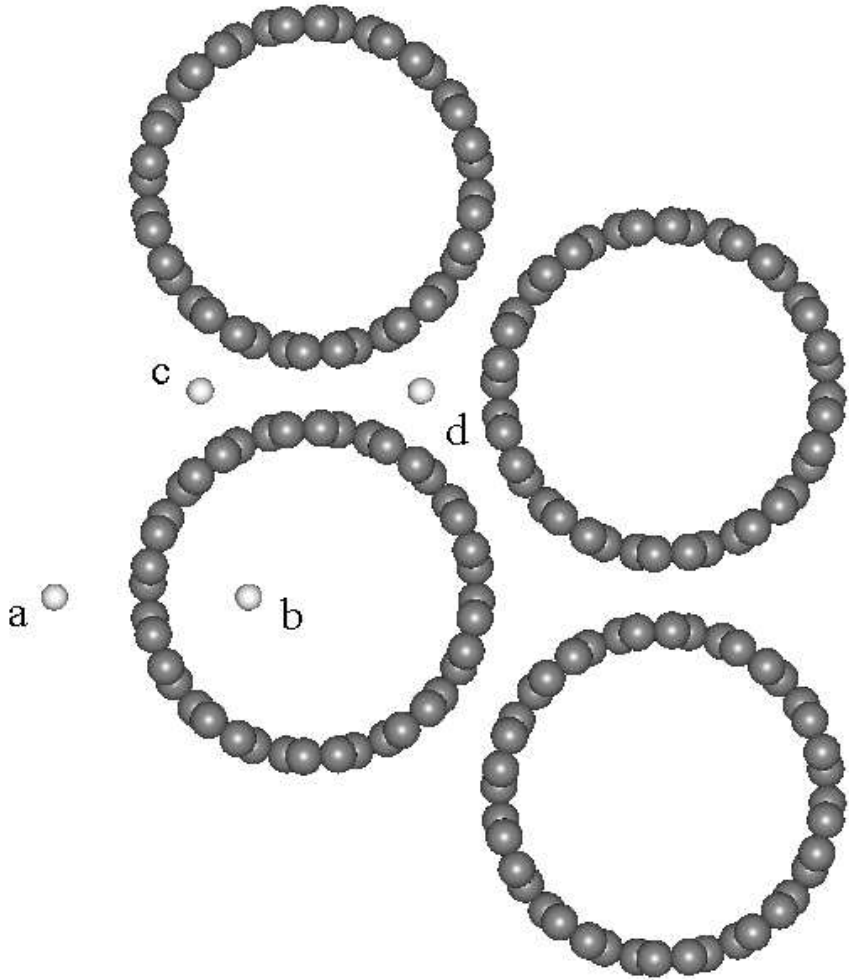}
}
\caption{Illustration of several adsorption sites for the H$_2$ molecule in the nanotube bundle. a: surface; b: pore; c: groove; d: interstial.}
\end{figure}

As an example of gas adsorption in the carbon nanotube bundle, we discuss the H$_2$ molecule adsorption in (10,10) SWNT bundle. As shown in Fig.3, four possible sites (surface, pore, groove, interstial) for the H$_2$ adsorption in the tube bundle have been considered. The calculated tube-molecule distance, adsorption energy and charge transfer for the different sites are given in Table III. We find that the adsorption energy and charge transfer of H$_2$ in the interstitial and the groove sites of tube bundle are considerably larger than those on the surface site. The pore site is also energetically more favorable than the surface site. Similar results are obtained for the other gas molecules studied. The enhancement of molecule adsorption on groove and interstitial sites can be understood by the increased number of carbon nanotubes interacting with molecule. Our present results compared well with a previous empirical force field simulation \cite{10}. 

\begin{table}
Table III. Equilibrium tube-molecule distance ($d$), adsorption energy ($E_a$) and charge transfer ($Q$) of H$_2$ molecule on different adsorption sites (see Fig.3) in (10,10) SWNT bundle.
\begin{center}
\begin{tabular}{cccc}
 site      & $d$ (\AA) & $E_a$ (meV)  &  $Q$(e)  \\ \hline
surface    &  3.01     &   94         &   0.014       \\
pore       &  2.83     &  111         &   0.012       \\
groove     &  3.33     &  114         &   0.026       \\
interstial &  3.33     &  174         &   0.035       \\
\end{tabular}
\end{center}
\end{table}

\subsection{Electronic properties of SWNT upon molecule adsorption}

The electronic band structures near the Fermi level for a (10,0) nanotube adsorbed with NH$_3$ and NO$_2$ are presented in Fig.4 and compared with that of pure nanotube. We find the band structures for either valence or conduction bands of carbon nanotube are not significantly changed upon the molecule adsorption. However, the degeneracies of the energy bands in SWNT are removed by the molecule-tube interaction. The band splitting cause by NO$_2$ is more pronounced than that by NH$_3$. This is consistent with the larger charge transfer and the higher adsorption energy (Table II). 

\ \\
\ \\
\begin{figure}
\centerline{
\epsfxsize=4.5in \epsfbox{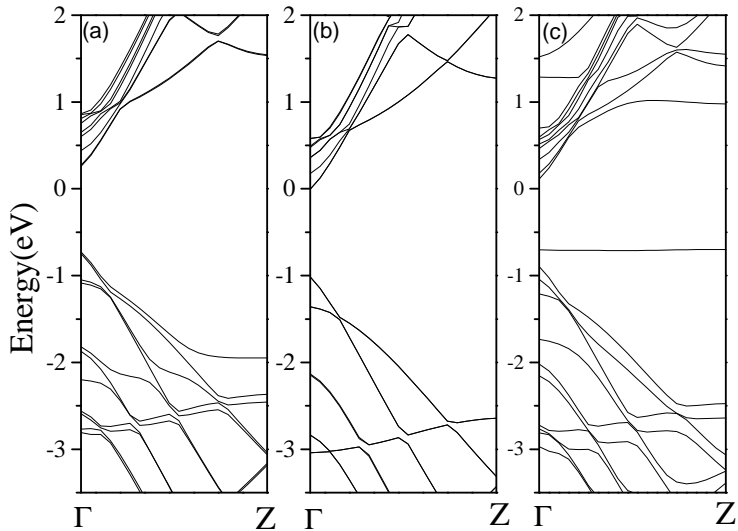}
}
\caption{Electronic band structures near Fermi level of (a) NH$_3$ adsorbed, (b) pure, (c) NO$_2$ adsorbed (10,0) carbon nanotube. The band degeneracies in SWNT are broke by molecule adsorption. For NO$_2$, this effect is enhanced by the presence of a molecule level near the tube valence band edge (see also Fig.6).}
\end{figure}

In Fig.5, we present the electronic density of states (DOS) of individual (10,0) SWNTs adsorbed with NO$_2$, NH$_3$, along with that of the pure SWNT. Except for the slight modification on DOS shape due to band splitting, we find the DOS of NH$_3$ adsorbed nanotube is very close to that of pure tube. Similar behavior is obtained for all charge donor molecules studied (N$_2$, H$_2$O, CO$_2$ etc.). Therefore, we suggest that the interaction between nanotube and these gas molecules is weak and does not have significant influence on the electronic structures of SWNTs. 

In the cases of NO$_2$ and O$_2$, the interaction between molecule and SWNT are much more pronounced. As shown in Fig.5, the adsorption of NO$_2$ has considerably modified the shape of DOS. Due to existence of a half occupied molecule level near the valence band edge, the coupling between tube and molecule shift the Fermi energy into the valence band and the DOS at the Fermi energy is high. Therefore, a semiconductor SWNT can be turned into $p$-type conductor after NO$_2$ or O$_2$ adsorption. Similar results have been obtained in a previous theoretical study on O$_2$ adsorption \cite{20}.

\begin{figure}
\centerline{
\epsfxsize=4.5in \epsfbox{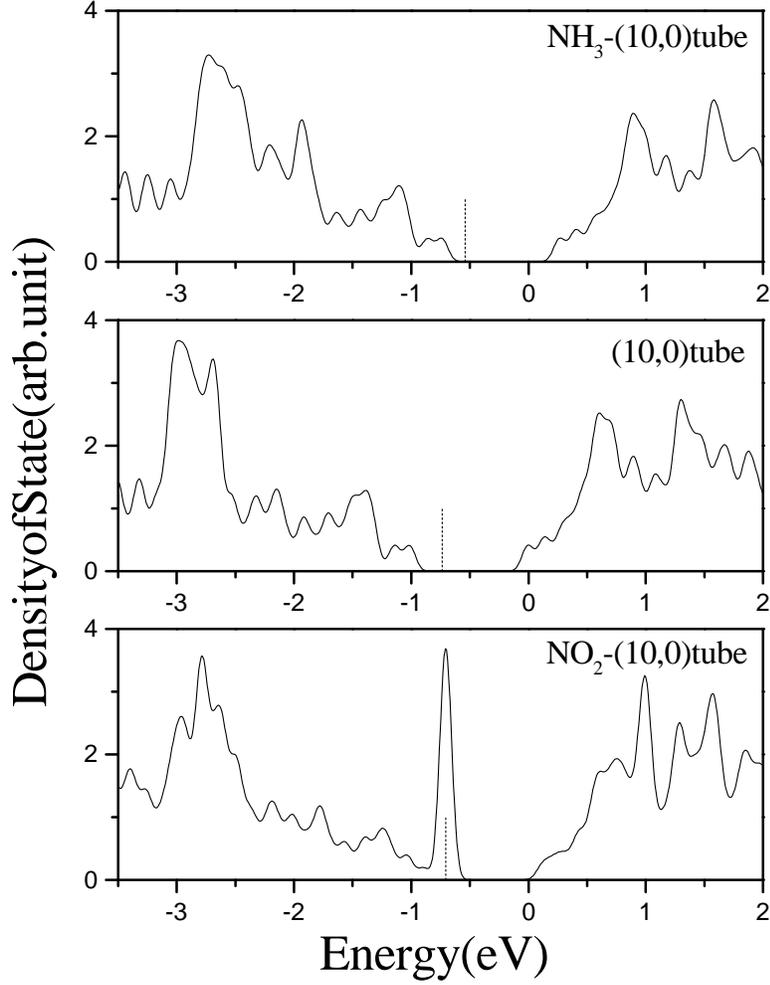}
}
\caption{Electronic density of states corresponding to the band structures shown in Fig.4. Dashed lines denote Fermi level. 21 {\bf k} points are used with 0.05 eV Guassian broadening. Modification of DOS caused by band splitting is more pronounced in NO$_2$ than NH$_3$. Due to existence of molecule level near the valence band edge, hybridization between SWNT and NO$_2$ shift the Fermi level into valence band, making SWNT $p$-type conductor. }
\end{figure}

Our present results are supported by recent experiments. For instance, the electrical conductance of an individual semiconducting tube increases dramatically upon NO$_2$ gas exposure and the NO$_2$ is identified as charge acceptor \cite{14}. Collins {\em et al} found that the oxygen gas has dramatic effects on conductivity, thermopower, and local density of states of individual semiconductor nanotubes, while Ar, He, and N$_2$ have no noticeable doping effect. Oxygen exposure generally converts semiconducting tubes into apparent conductors \cite{15}. NMR experiment has further proved the increase of density of state at Fermi level of SWNTs after exposure to oxygen \cite{16} while most other gas like H$_2$, CO$_2$ do not have such effect \cite{28}. Our calculations suggest that this is probably due to the hybridization between the tube valence bands and nearby molecule level. As suggested in Ref.[15], the air exposure effect on the measured properties of as-prepared nanotubes should be carefully examined. Many supposedly intrinsic properties measured on tube samples might be severely compromised by extrinsic air exposure effects. Based on present calculations, we propose that the effects from most gas molecules in the air, such as N$_2$, CO$_2$, H$_2$O, are relatively weak. The air exposure effect should be dominated by O$_2$, which is charge acceptor and make all nanotubes $p$ type conductors. 

\begin{figure}
\centerline{
\epsfxsize=4.5in \epsfbox{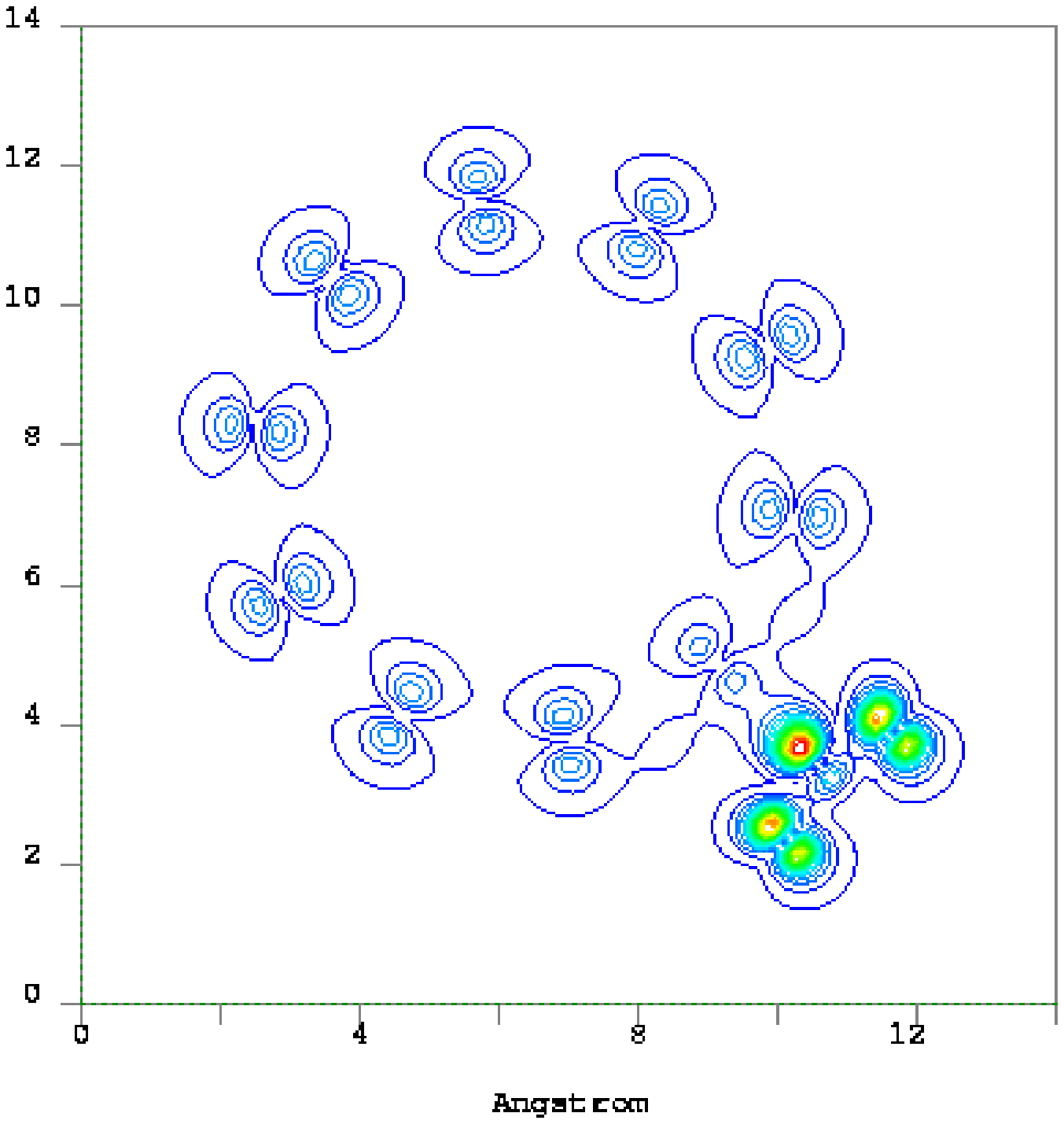}
}
\caption{Contour plot of electron density (slice at (100) direction and passing through NO$_2$ molecule) for the top nine valence bands. The corresponding band structures of these bands are shown in Fig.5.}
\end{figure}

More insight into tube-molecule interaction can be obtained by examining the electron density. Shown in Fig.6 is the electron density of top nine electronic bands near the valence band edge (as shown in Fig.5) of NO$_2$ adsorbed (10,0) tube. Weak coupling is found between NO$_2$ and carbon nanotube. The carbon $\pi$ bonds near the molecule are weakened due to the charge transferred from carbon to NO$_2$. We also find charge density redistribution on carbon close to the NO$_2$ molecule. Therefore, we think that the molecule adsorption can induce a local charge fluctuation in the contact region of the nanotube. Such fluctuation may have pronounced effect on the transport properties of metallic SWNT. The charge fluctuation can act as scatting centers for conducting electrons and increase the resistance of a metallic nanotube. In a recent experiment, it was found that the resistance of metallic tube bundle increase with gas molecules (O$_2$, N$_2$, etc) \cite{17}. We have carried out preliminary resistance calculations on the molecule induced charge fluctuations found that significant resistance increase can be detected with a single molecule adsorption. The detailed study on this subject will be reported elsewhere \cite{29}.

\section{Conclusions}

In summary, we have performed first principles calculations on the electronic properties of nanotube upon adsorption of gas molecules. We found that all molecules are weakly adsorbed on SWNT with small charge transfer, while they can be either charge donor or accepter of nanotube. The adsorption of some gas molecules on SWNTs can cause significant change in electronic and transport properties of nanotube due to the charge transfer and charge fluctuation. The molecule adsorption on the surface or inside of the nanotube bundle is stronger than that on an individual tube. 

\begin{acknowledgements}
This work is supported by the U.S. Army Research Office Grant No. DAAG55-98-1-0298, the Office of Naval Research Grant No. N00014-98-1-0597 and NASA Ames Research Center. The authors thank Dr.X.P.Tang, Dr.A.Kleinhammes and Prof. Y.Wu for helpful discussions. We acknowledge computational support from the North Carolina Supercomputer Center.
\end{acknowledgements}

\ \\
$^*$: zhaoj@physics.unc.edu \\
$^{\dagger}$: jpl@physics.unc.edu


\begin{references}
\bibitem{27} Tube-molecule distance $d$ is defined as the nearest distance between atoms on molecule and nanotube for T site, or the distance between the center of gas molecule and the center of carbon hexagon (carbon-carbon bond) for C (B) site. The adsorption energy $E_a(d)$ is defined as the total energy gained by molecule adsorption at equilibrium distance: $E_a(d)=E_{tot}(tube+molecule)-E_{tot}(tube)-E_{tot}(molecule)$. Charge transfer $Q$ denotes the total Mulliken charge number on the molecules, positive Q means charge transfer from molecule to tube.
\end{references}
\end{document}